# Measurement of airborne $^{131}$I, $^{134}$Cs, and $^{137}$Cs nuclides due to the Fukushima reactors accident in air particulate in Milan (Italy)

M. Clemenza, E. Fiorini, E. Previtali, E. Sala,

$^a$University of Milano Bicocca Physics Department "G. Occhialini" and INFN sec. of Milano Bicocca - Piazza della Scienza 3, 20126 Italy

**Abstract.** After the earthquake and the tsunami occurred in Japan on 11$^{th}$ March 2011, four of the Fukushima reactors had released in air a large amount of radioactive isotopes that had been diffused all over the world. The presence of airborne $^{131}$I, $^{134}$Cs, and $^{137}$Cs in air particulate due to this accident has been detected and measured in the Low Radioactivity Laboratory operating in the Department of Environmental Sciences of the University of Milano-Bicocca. The sensitivity of the detecting apparatus is of 0.2 µBq/m$^3$ of air. Concentration and time distribution of these radionuclides were determined and some correlations with the original reactor releases were found. Radioactive contaminations ranging from a few to 400 µBq/m$^3$ for the $^{131}$I and of a few tens of µBq/m$^3$ for the $^{137}$Cs and $^{134}$Cs have been detected

**Keywords:** Airborne radioactivity, Fukushima accident, $^{131}$I, $^{137}$Cs and $^{134}$Cs

## INTRODUCTION

The laboratory for low radioactivity measurements of the Physics department of the University of Milano Bicocca and of the Italian Institute of Nuclear Physics INFN has been operating for fifteen years in the basement of the Environmental Sciences Department. The main scope of this laboratory is material selection and study of very low radioactive contaminations as contribution to the background in searches on rare events. This is presently in view of the CUORE experiment [1] now under construction in the *Laboratori Nazionali del Gran Sasso* of INFN. For such type of applications an extreme sensitivity in the determination of very low radioactive contaminations is mandatory. As a by-product, thanks to its sensitivity, this facility was used also for other scientific topics like: environmental measurements, historical and archaeological characterization and other studies not directly connected with experiments in nuclear and particle physics. Many measurements on air pollution in the Milan area were carried out during the last 15 years. Tests on $^{222}$Rn concentration in air were routinely performed in order to trace the particulate evolution, such as the presence of $^7$Be for the analysis of the PBL (Planetary Boundary Layer). In the case of $^{137}$Cs the presence of a tiny contamination of about 30 µBq/m$^3$ was detected, due to the accidentally release in Algeciras (Spain) at the end of May 1998; the contamination was rapidly reduced to an almost constant value of 0.5-1 (µBq/m$^3$) which is the background value now measured in Milan's air [2].

## EXPERIMENTAL METHOD AND PROCEDURES

The detection limit of these measurements is related to the total volume of the filtered air, to the particulate retention efficiency of the filters and to the absolute gamma ray detection efficiency of the High Purity Germanium (HPGe) detector used for gamma rays spectroscopy. A ventilation system made by 9 independent filters, Class G4, of polypropylene fibers operating in the roof of the Environmental Science Department of the University of Milano Bicocca was used to collect the air particulate. The filters have a dimension of 397x498x48 mm$^3$ with a air flux of 1700 m$^3$/h for each filter and a retention capability for air particulate of more than 95% for particulate larger than 0.4 μm. The position inside the ventilation system and the time exposure of each filter were controlled each day in order to check the integral volume fluxed by the system.

The HPGe detector selected for these measurements are low background ones with a relative detection efficiency of about 30%. The energy resolution at the 1,3 MeV gamma lines of $^{60}$Co is 1.8 keV FWHM.

Each filter was removed and changed every day: it was dismounted from the metallic frames and the polypropylene parts were rolled in a polyethylene film and then placed all around the end cap of HPGe detector (Fig. 1a, 1b ) in a way to maximize the absolute detection efficiency for the emitted gamma lines.

Measurements were normally subdivided in two different period: the first one shows a background dominated by the $^{222}$Rn and $^{220}$Rn daughters decay; a second period in which the background correlated with the analyzed sample is low making possible a more sensitive analysis of the radioactive contaminations. Total measurement time was normally of about 24 hours just to allow the change of one filter a day. Sometimes the measured filter was placed on a second HPGe detector with the same characteristics as the first one, to look for a possible long running time measurements to better determine long living isotopes.

The complete sequence of measurement was carried out for around 40 days starting on March 15. In order to increase the sensitivity and to look for a possible presence in air particulate of other less abundant radionuclides like $^{136}$Cs, we have removed at the end of the sequence 5 filters that had remained in place all the time period in which the Fukushima radioactive clouds were in the air of Milan area.

Montecarlo simulations with Geant4 toolkit were done in order to correctly evaluate the HPGe detection efficiency [3] taking into account the relative position of filters with respect to the detector. The complete decay scheme of the analyzed nuclei was reconstructed in order to take into account all possible aspects of the measured spectra, especially those connected with the emission of gamma cascade that can change the efficiency evaluation for some nuclei. In particular the decay schemes of $^{131}$I, $^{137}$Cs, $^{134}$Cs and $^{136}$Cs shown in Fig. 2a,b,c,d), were taken into account in the simulation analysis.

## RESULTS

The air in Milan area was monitored also before the Fukushima accident, but we will discuss in this analysis the measurement sequence as started on March 15, when no effect of the Fukushima radioactive clouds was expected in Northern Italy. Using this data we were able to fix a reference of the radioactive background normally present in the air pollution of the Milan area just before the arrival of the radioactive clouds, as shown in Fig. 3. The presence of the $^{222}$Rn and $^{220}$Rn daughter is clearly evident, with the counting rate of the former decreasing during the first couple of hours. There are also a dominant peaks due to $^{40}$K caused by the re-suspension of dust from soils and a $^{7}$Be peak from cosmogenic activations in air. This latter is particularly evident due to the season, spring, when the PBL over the Milan sky is particularly thin.

To evaluate possible interference with the analysis of the anthropogenic radioactivity released by the Fukushima power plant, we have previously analyzed the presence of nuclei released in the environment by human activities: specifically those related to nuclear test explosion in the atmosphere and to the Chernobyl accident. In this analysis we found that only a residual of $^{137}$Cs was present, with a contamination in the air ranging from 0.5 to 1 μBq/m$^3$. This contamination had been monitored for a long time and no evidence of large fluctuation in air was found, apart the previously mentioned one due to the release of this isotope from Spain. This result indicates that our measurement sensitivity is in the scale of less than 1 μBq/m$^3$ and the stability of these data demonstrates that systematic errors are under control and that the change of the filters doesn't affect the results. It give us also the precious information that a real background due to $^{137}$Cs can be expected in the measurement of the Fukushima radioactive clouds.

Starting from March 24 our instrumentation detected evidence of radioactive nuclei connected with the release from nuclear reactors. In particular $^{131}$I, $^{137}$Cs and $^{134}$Cs were identified. In our survey the first interesting evidence was $^{131}$I that was detected in advance by about two days with respect to the Cs isotopes. It is not clear if this delay can be attributed to different releases from the Fukushima power plants or to other not identifiable reasons. In Fig. 4 the gamma lines emitted by the various isotopes are clearly identifiable and we can reconstruct the concentration in air of these radioactive contaminants. It is also possible to reconstruct the time variation, day by day, of the pollution. The presence of the radioactive cloud in the north of Italy was detected for approximately 22 days.

The contaminations of $^{131}$I , $^{134}$Cs and $^{137}$Cs are reported as function of time in Fig. 5. It is important to note that, in our analysis, the time distribution of the Cs concentration in air was the same for the two isotopes, $^{137}$Cs and $^{134}$Cs, as expected, with a $^{137}$Cs/$^{134}$Cs ratio close to one. This ratio is in perfect agreement with that measured in Japan by KEK Collaboration at Tsukuba [4] and by Fushimi et al. at Tokoshima [5]; $^{134}$Cs is not a direct fission product but is produced by neutron activation of $^{133}$Cs, which is the decay product of $^{133}$Xe present in nuclear reactors [6]. The ratio between the concentration of the two Cs isotopes measured in Italy and in Japan is the same demonstrating that the measurements are referred at the same event. Obviously the elements concentrations recorded in Milan are lower from those detected at distance of 165 km (Tsukuba) and 700 km (Tokoshima) from Fukushima. The measurement of 5 filters together (from 15$^{th}$ March to 11$^{th}$ April) allows the detection of a larger air flux improving the counting statistics; in this way it was possible to estimate the mean value of $^{136}$Cs contamination in air during the period in which the Fukushima radioactive clouds were in Northern Italy. The mean contamination of this isotope was evaluated in $4,6 \pm 1,4$ nBq/m$^3$.

Our data demonstrate that there is no evidence of any possible sanitary impact on the resident population of our region related with the Fukushima radioactive release. At the end of the period of our measurement the radioactive pollution in air was found to return approximately at the same levels as before the arrival of the Fukushima radioactive clouds.

## CONCLUSIONS

A contamination of $^{131}$I, $^{134}$Cs, and $^{137}$Cs ranging from a few up to a few hundred µBq/m$^3$ has been measured in air particulate of Milan as a consequence of the incident at the Fukushima nuclear reactor facility. This is confirmed by measurements carried out in the rest of Italy [7], Europe [8], [9] and USA[10]. The correlation between the event at Fukushima and the data recorded at Milan is supported by the agreement in the ratio $^{137}$Cs/$^{134}$Cs.

## ACKNOWLEDGMENTS

Our sincere thanks to the graduate D. Rigamonti for the help in the analyses of the spectra and to INFN for the financial support.

# FIGURES AND TABLES

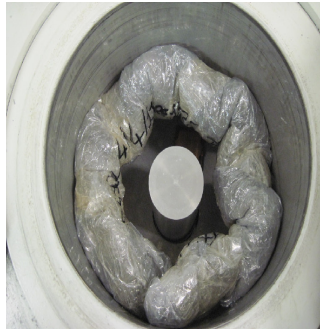

**FIGURE 1a).** Filter rolled in a polyethylene film and placed all around the end cap of HPGe detector.

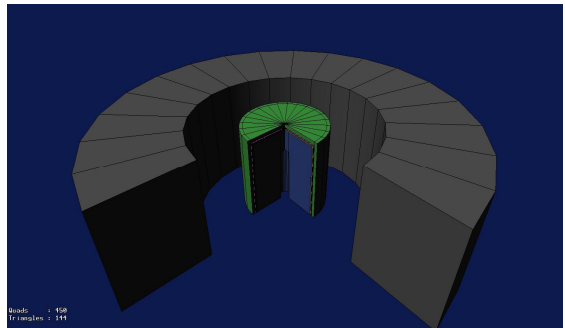
**FIGURE 2b).** Monte Carlo Simulation

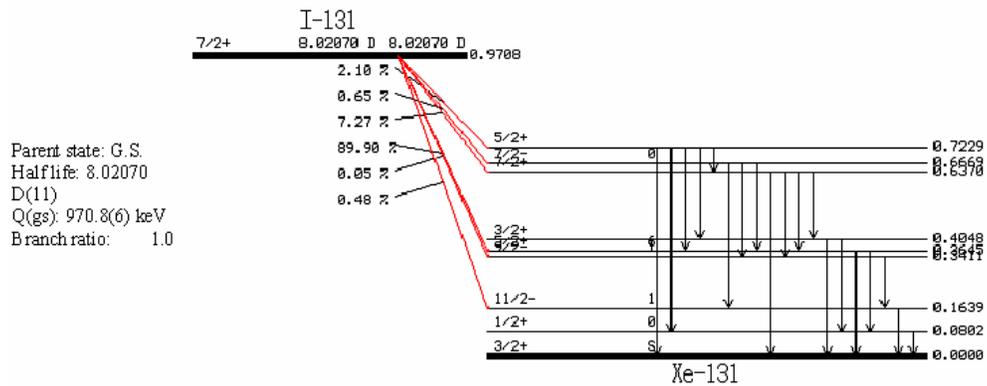
**FIGURE 2a.** Decay scheme of $^{131}$I (ref. http://atom.kaeri.re.kr/)

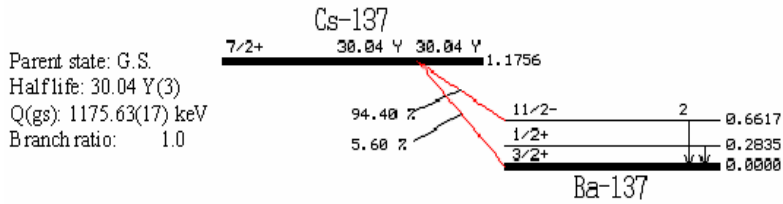

**FIGURE 2b.** Decay scheme of $^{137}$Cs (ref. http://atom.kaeri.re.kr/)

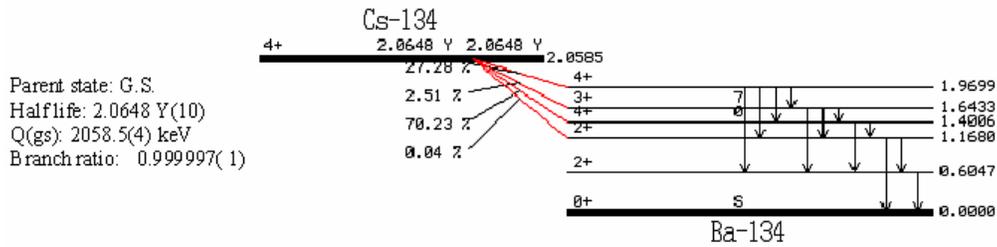

**FIGURE 2c.** Decay scheme of $^{134}$Cs (ref. http://atom.kaeri.re.kr/)

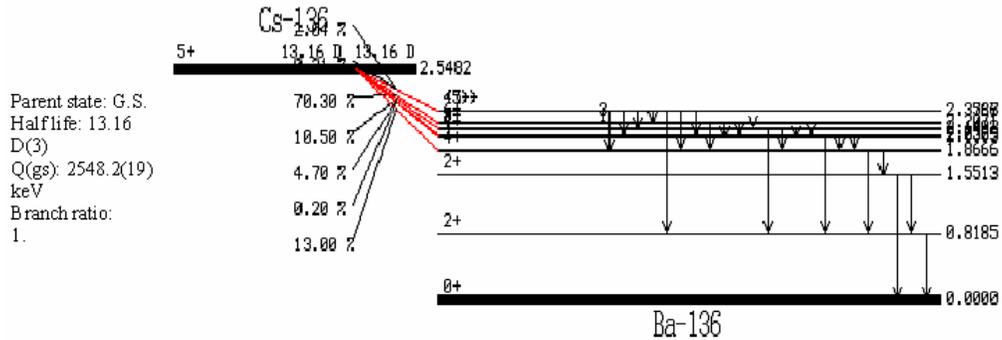

**FIGURE 2d.** Decay scheme of $^{136}$Cs (ref. http://atom.kaeri.re.kr/)

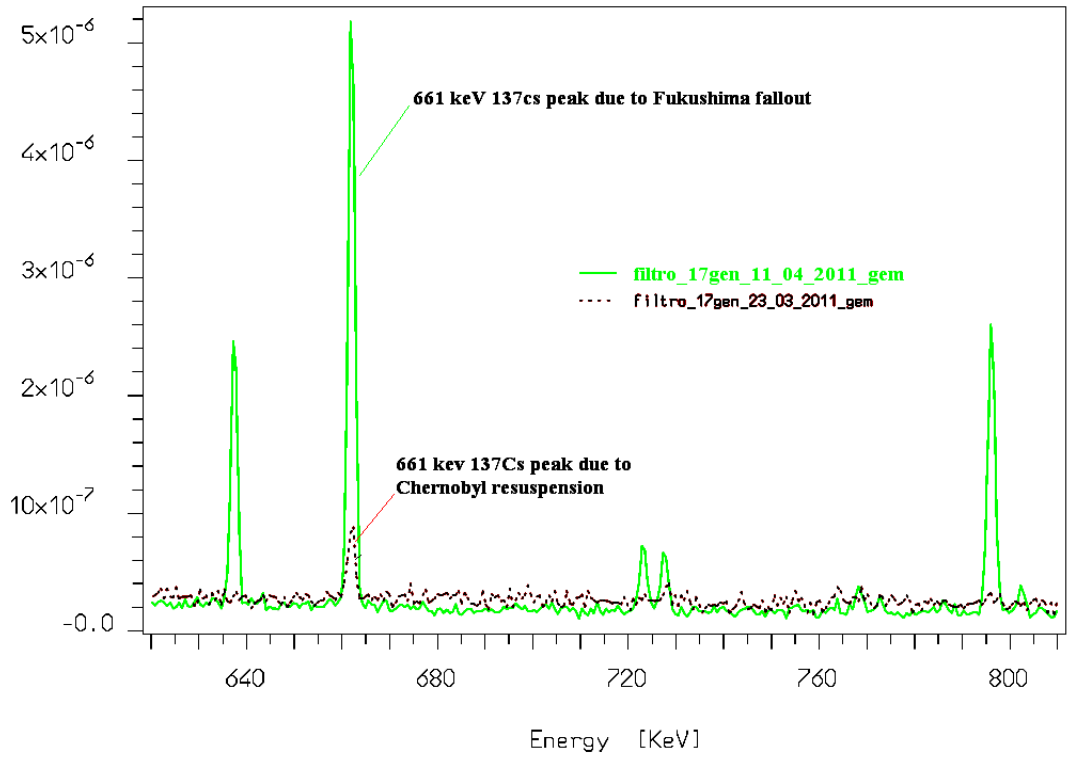

**FIGURE 3.** $^{137}$Cs peaks due to Chernobyl particulate re-suspension and Fukushima fallout

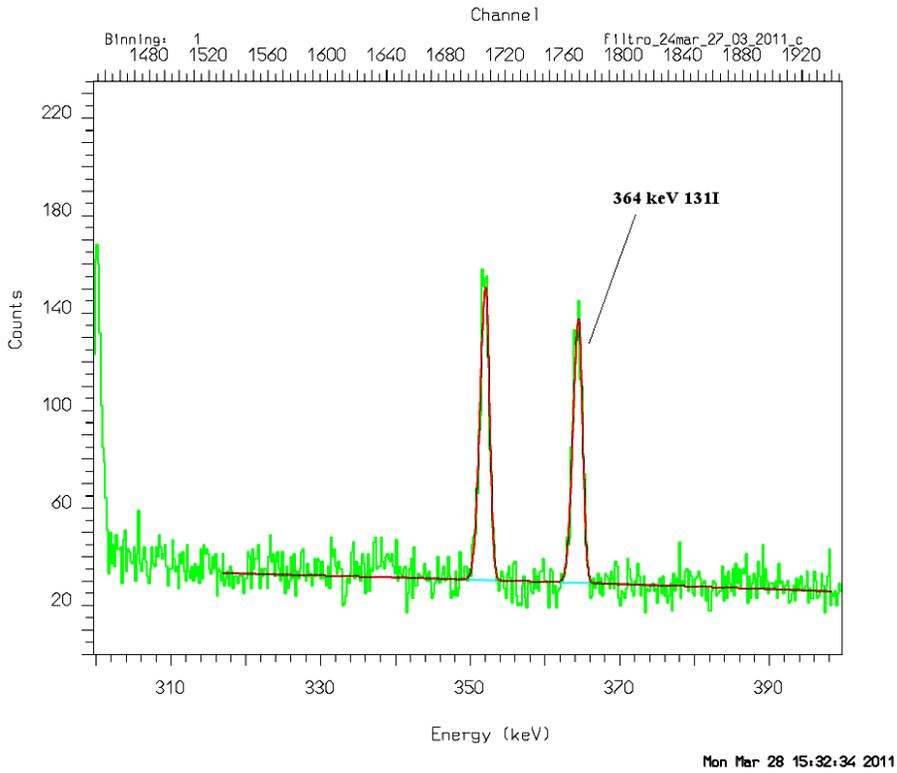

**FIGURE 4.** Gamma lines 1filter

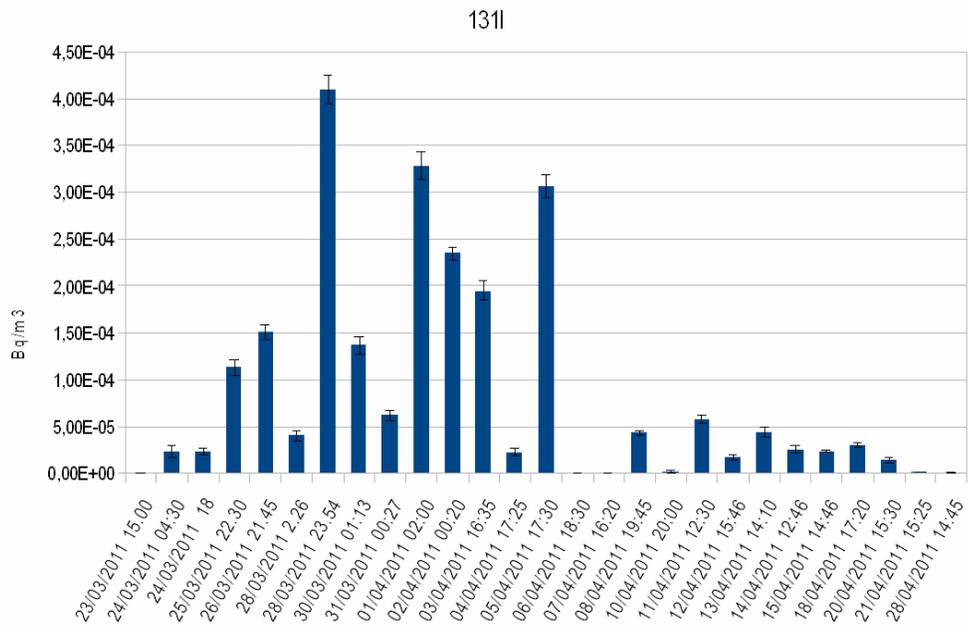

**FIGURE 5a.** Daily concentration in air particulate of $^{131}$I

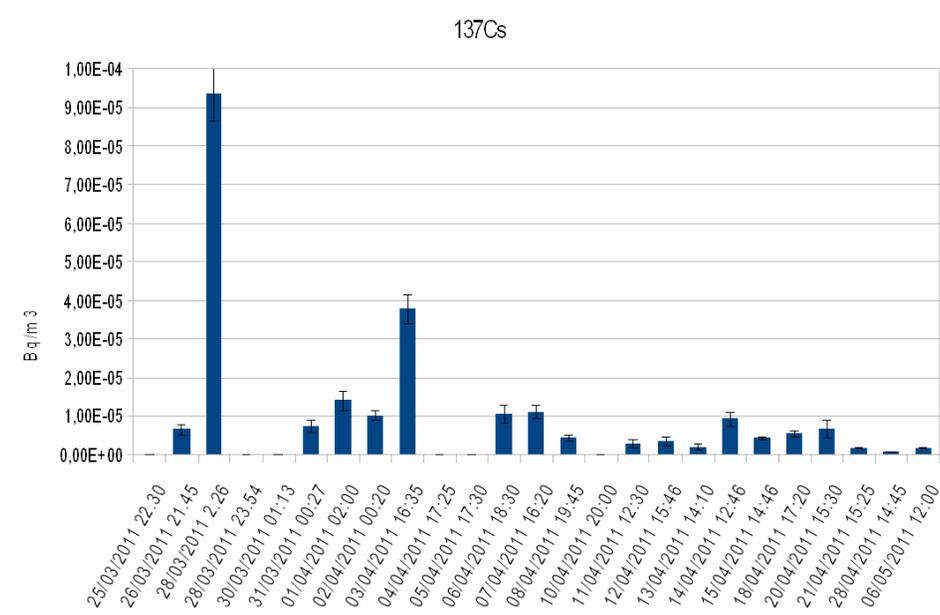

**FIGURE 5b.** Daily concentration in air particulate of $^{137}$Cs

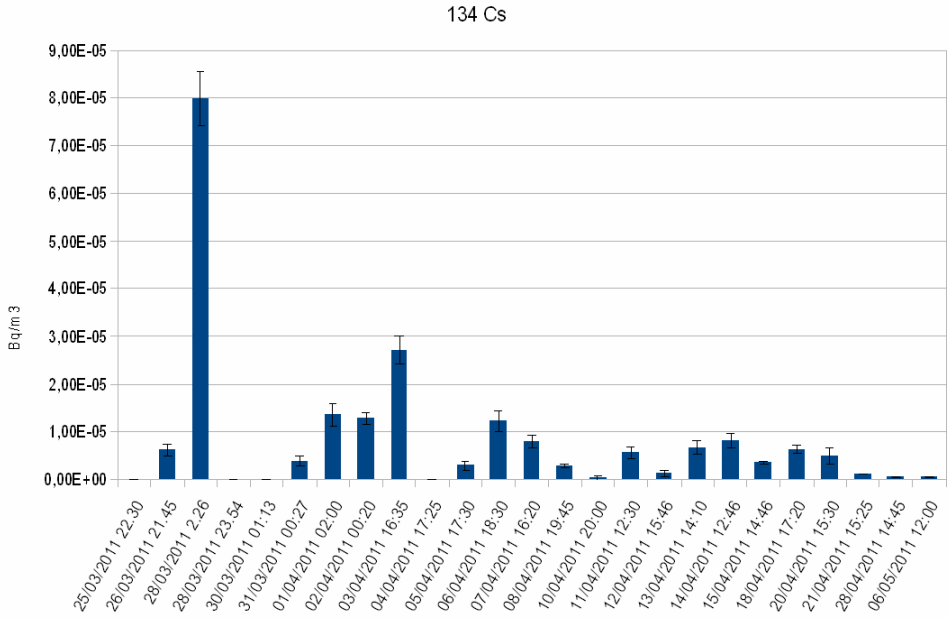

**FIGURE 5c.** Daily concentration in air particulate of $^{134}$Cs

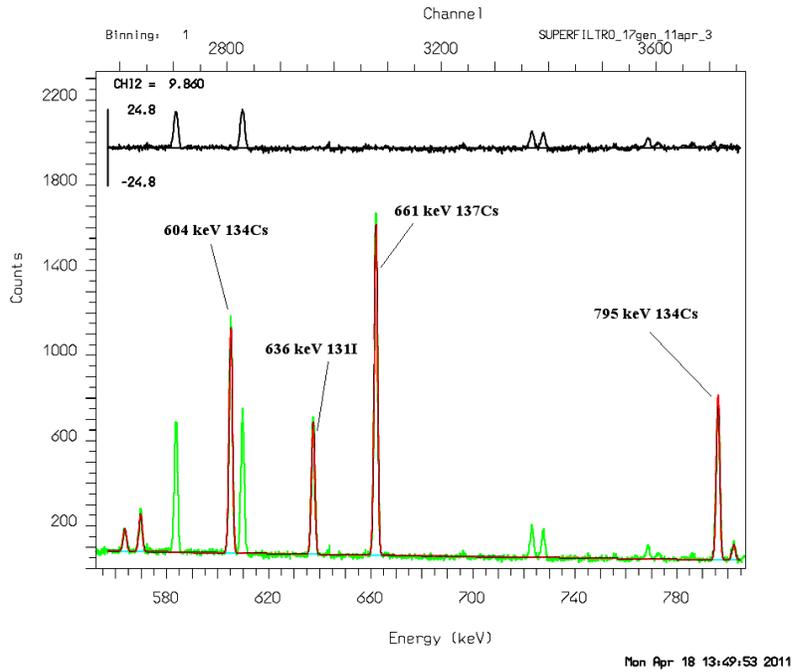

**FIGURE 6.** Spectra of 5 filters